\documentclass[12pt]{iopart}

\usepackage{graphicx}

\begin{document}

\title[Trigonometry and kinematics]{Correspondence between geometrical and differential definitions of the sine and cosine functions and connection with kinematics}

\author{Horia I. Petrache}

\address{Indiana University Purdue University Indianapolis, \\ Department of Physics, Indianapolis, IN 46202, U.S.A.}
\ead{hpetrach@iupui.edu}
\begin{abstract}
In classical physics, the familiar sine and cosine functions appear in two forms: (1) geometrical, in the treatment of vectors such as forces and velocities, and (2) differential, as solutions of oscillation and wave equations. These two forms correspond to two different definitions of trigonometric functions, one geometrical using right triangles and unit circles, and the other employing differential equations. Although the two definitions must be equivalent, this equivalence is not demonstrated in textbooks. In this manuscript, the equivalence between the geometrical and the differential definition is presented assuming no a priori knowledge of the properties of sine and cosine functions. We start with the usual length projections on the unit circle and use elementary geometry and elementary calculus to arrive to harmonic differential equations. This more general and abstract treatment not only reveals the equivalence of the two definitions but also provides an instructive perspective on circular and harmonic motion as studied in kinematics. This exercise can help develop an appreciation of abstract thinking in physics.  
\end{abstract}

\today
\maketitle

\section{Introduction}
For a number of years, I have surprised my physics students (in both introductory and graduate physics classes) with a "simple" question: What is sine? Or in extended form: How do we define the sine function that appears everywhere in physics? Such questions usually elicit amusement, perplexity, and entertaining statements such as "The sine function need not be defined because it is well known."  Discussions that follow the initial puzzled reactions are quite illuminating and inevitably lead to the question of what properties of trigonometric functions can be easily derived from others without prior knowledge. For example, if we start with the geometrical definition of the sine and cosine functions that uses right triangles (as it appears in vector analysis), can we show that the function defined this way has the property that $f'' = -f$, as in harmonic motion? This manuscript addresses the equivalence between the geometrical and the differential definitions, first mathematically, and then in connection with circular and harmonic motion. This approach can help students enhance their ability to think abstractly in addition to acquiring more insight into trigonometric functions.

\section{Derivations on the unit circle}
Consider a unit circle as illustrated in Fig. 1(a) which shows the projection of the radius OP on the $x$ and $y$ axes. These projections are functions of the arc length $s$. Assuming no apriori knowledge on trigonometric functions, we would like to show that the functions $x(s)$ and $y(s)$ are harmonic, meaning that $x'' = -x$ and $y'' = -y$, where the derivative is taken with respect to $s$. 

Before doing that, let's ask what can be said about $x(s)$ and $y(s)$, as defined on the unit circle. The most obvious property is the circle equation 
\begin{equation}
x^2 + y^2 = 1.
\label{eq:sq1}
\end{equation}

Does this equation alone specify the sine and the cosine? The answer is no, although a number of familiar properties follow immediately from it, such as (1) both functions are bounded with values between -1 and 1, and (2) when $x$ is maximum or minimum, $y$ is zero, and vice versa. To specify the sine and the cosine, a second relationship is needed as shown below. This second relationship is obtained geometrically, (almost) like the first.

\subsection{From geometry to differential equations}

Consider a small displacement from P to P'  over a distance $ds$ along the unit circle, as shown in Fig. 1(b). The $x$ and $y$ projections change by $dx$ and $dy$, respectively.  We have
\begin{equation}
(dx)^2 + (dy)^2 = (ds)^2
\end{equation}
Divide by $(ds)^2$ to get
\begin{equation}
(dx/ds)^2 + (dy/ds)^2 = 1,
\end{equation}
or in short notation,
\begin{equation}
x'^2 + y'^2 = 1.
\label{eq:sq2}
\end{equation}
We can now manipulate Eqs.~\ref{eq:sq1} and \ref{eq:sq2} to obtain $x'' = -x$ as follows. Differentiate  Eq.~\ref{eq:sq1} to obtain
\begin{equation}
xx' = - yy'.
\label{eq:xx}
\end{equation}
Squaring the last equation gives $x^2x'^2 = y^2 y'^2$ in which $y^2$ and $y'^2$ can be substituted from Eq.~\ref{eq:sq1} and Eq.~\ref{eq:sq2}, respectively. We obtain
\begin{equation}
x^2x'^2 = (1-x^2)(1-x'^2),
\end{equation}
which reduces to 
\begin{equation}
x^2+x'^2=1.
\label{eq:xsq}
\end{equation}
Differentiating this last equation gives
\begin{equation}
x'(x+x'')=0,
\end{equation}
and since $x'$ is not identically zero, it follows that $x+x''=0$. 
The corresponding equation for $y$ is obtained in a similar fashion.

Note that the result in Eq.~\ref{eq:xsq} implies that $x' = \pm y$ and $y' = \mp x$, showing that the $x$ and $y$ functions are derivatives of each other as expected. 
Furthermore, from $x'' = -x$ and $y'' = -y$, we obtain 
\begin{equation}
x''^2+y''^2 =  x^2 + y^2 = 1,
\label{eq:acc}
\end{equation}
and in general $(x^{(n)})^2 + (y^{(n)})^2 = 1$, where $x^{(n)}$ and $y^{(n)}$ represent the $n$-derivatives of $x$ and $y$ with respect to $s$.

\subsection{From differential equations to geometry}

We consider two real functions $x(s)$ and $y(s)$ with the following two properties:
\begin{eqnarray}
x'' &=& -x \\
y  &=& \pm x' 
\label{eq:harmonic}
\end{eqnarray}
and ask whether a convenient geometrical representation of $x$ and $y$ exist (other than a plot vs. the variable $s$). By direct differentiation, it can be shown that $x^2 + y^2 = const.$ and $x'^2 + y'^2 = const.$ which brings us back to the unit circle as a possible representation of $x$ and $y$ by choosing the constants to be equal to 1.

\section{Connection with physics and further reading}

In the above development, we have used the variable $s$ to represent the length of arcs along the unit circle. If we consider an object moving around the circle with constant speed $v$, then the geometrical variable $s$ is proportional to physical time $t$ through the speed $v$, 
\begin{equation}
s = v t,
\end{equation}
and 
\begin{equation}
ds = v dt.
\end{equation}

Using the standard dot notation for derivatives with respect to $t$, we have 
\begin{eqnarray}
\dot{x} &=& v x' \\
\dot{y} &=& v y'. 
\end{eqnarray}
Note that $\dot{x}$ and $\dot{y}$ have units of speed as expected, while $x'$ and $y'$ are dimensionless.

Eq.~\ref{eq:sq2} then expresses the fact that ${\dot{x}}^2 + {\dot{y}}^2 = v^2$. In addition, using the second derivatives of $x$ and $y$, we should recover the known expression $a = v^2/R$ for acceleration. To do this, we need to reintroduce the radius of the circle into equations. Following the same derivation steps as above, it can be shown that
\begin{eqnarray}
x^2 + y^2 & = & R^2 \\
x'^2 + y'^2 & = &	1 \\
x''^2 + y''^2 & = & R^{-2}.
\end{eqnarray}
The last relationship follows from  Eq.~\ref{eq:xsq} which becomes $x^2 + R^2 x'^2 = 1$ giving 
$x'' = -x/R^2$ and  $y'' = -y/R^2$. 
It follows that
\begin{equation}
\ddot{x}^2+\ddot{y}^2 = v^4(x''^2 + y''^2) = \frac{v^4}{R^2}
\end{equation}
so that the magnitude of the (centripetal) acceleration is 
\begin{equation}
a =  \sqrt{\ddot{x}^2+\ddot{y}^2} = \frac{v^2}{R}.
\end{equation}

A number of alternative and instructive derivations of the centripetal acceleration can be found in references 3-8.
  
Why bother with derivations that make no explicit use of trigonometric functions? After all, trigonometric functions are well known and well established and allow for more elegant derivations. There are at least two different perspectives that are worth considering. From a mathematical perspective, the question of possible characterizations of sine and cosine functions has received significant interest. In this context, characterization of a function means finding a property that is obeyed only by that function. Different characterizations of the same function must therefore be equivalent. Interested readers are directed to more advanced material in references 9-12. From a physics perspective (in kinematics to be more precise), we have seen that the connection between geometrical and differential definitions of sine and cosine functions corresponds to the familiar connection between circular and harmonic motions at a more fundamental level than commonly presented in textbooks. 
\vspace{.2in}

Acknowledgments

The author thanks Derek Scott, Andrew Seeran, Heather Stout, and Kashyap Vasavada for valuable discussions and suggestions for the manuscript.

\vspace{.2in}

\textbf{References}\\
$^1$A. R. Crathorne and E. B. Lytle, \textit{Trigonometry} (Henry Holt and Company, New York, 1938). \\
$^2$E. P. Vance, \textit{Trigonometry} (Addison-Wesley Publishing, Cambridge, 1954).\\
$^3$H. W. Jones and J. S. Wallingford, "A simple derivation of centripetal force," Am. J. Phys. 37, 751 (1969).\\
$^4$E. Zebrowski, Jr., "On the Derivation of the Centripetal Acceleration Formula,"  \textit{Phys. Teach.} 10, 527-528 (1972).\\
$^5$R. D. Patera, "Deriving the centripetal acceleration formula: $a = v^2/r$", Phys. Teach. 13, 547 (1975).\\
$^6$B. Wedemeyer, "Centripetal acceleration a simpler derivation,"  Phys. Teach. 31, 238 (1993).\\
$^7$F. Ninio, "Acceleration in uniform circular motion," Am. J. Phys. 61, 11 (1993).\\
$^8$H. S. Leff, "Acceleration for circular motion," Am. J. Phys. 70, 490--492 (2002).\\
$^9$J. Roe, "A characterization of the sine function," Math. Proc. Camb. Phil. Soc. 87, 69--73 (1980).\\
$^{10}$R. Howard, "A note on Roe's characterization of the sine function," Proc. Amer. Math. Soc. 105, 658--663 (1989).\\
$^{11}$L. J. Wallen, "One parameter groups and the characterization of the sine function," Proc. Amer. Math. Soc. 102, 59--60 (1988).\\
$^{12}$J.-H. Kim, Y.-S. Chung, S.-Y. Chung, "Generalization of characterizations of the trigonometric functions," Math. Proc. Camb. Phil. Soc. 141, 409-519 (2006).

\newpage

\begin{figure}[h!]
\centering
\includegraphics[width=6.0 in]{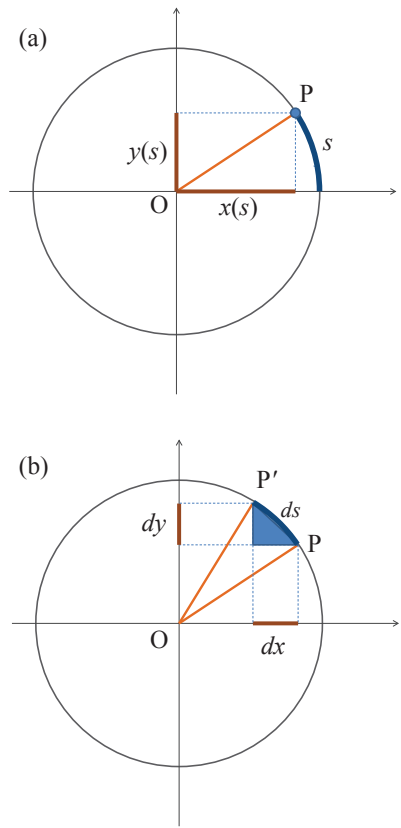}
\caption{(a) Length projections on $x$ and  $y$ axes define the cosine and the sine functions using the unit circle. (b) A small displacement $ds$ from P to P'  produces variations $dx$ and $dy$ in $x$ and $y$ projections. The displacement is exaggerated for clarity.}
\label{fig:sample}
\end{figure}

\end{document}